# Neutron inelastic scattering study of rare-earth orthoferrite HoFeO$_3$

A. K. Ovsyanikov[a,d], I. A. Zobkalo[a*], W. Schmidt[b], S.N. Barilo[c], S.A. Guretskii[c], V. Hutanu[d]

[a]*B.P Konstantinov Petersburg Nuclear Physics Institute, Kurchatov Institute, Gatchina, 188300, Russia;*
[b] *Julich Centre for Neutron Science Outstation at Institut Laue-Langevin, Grenoble, France;*
[c]*Scientific-Practical Materials Research Centre NAS of Belarus, 19 P. Brovki str., Minsk, 220072, Belarus;*
[d]*Institute of Crystallography, RWTH Aachen University and Jülich Centre for Neutron Science (JCNS) at Heinz Maier- Leibnitz Zentrum (MLZ), 85747 Garching, Germany.*

[*] Corresponding author: Igor A. Zobkalo, e-mail: zobkalo@pnpi.nrcki.ru, phone +7 (81371) 46416, address: Condensed Matter Research Department, PNPI, Orlova Roshcha, Gatchina, St. Petersburg distr., 188300, Russia.

**Abstract**

By the single crystal inelastic neutron scattering the orthoferrite HoFeO$_3$ was studied. We show that the spin dynamics of the Fe subsystem does not change through the spin-reorientation transitions. The observed spectrum of magnetic excitations was analyzed in the frames of linear spin-wave theory. Within this approach the antiferromagnetic exchange interactions of nearest neighbors and next nearest neighbors were obtained for Fe subsystem. Parameters of Dzyaloshinskii-Moriya interactions at Fe subsystem were refined. The temperature dependence of the gap in Fe spin-wave spectrum indicates the temperature evolution of the anisotropy parameters. The estimations for the values of Fe-Ho and Ho-Ho exchange interaction were made as well.

1. **Introduction**

The remarkable magnetic properties of rare-earth orthoferrites RFeO$_3$ result from complex interactions between the moments of 3*d* electrons of the transition metal and 4*f* electrons of the rare-earth. Investigations of these magnetic compounds were started several decades ago [1], including neutron powder diffraction investigations of the crystal and magnetic structures [2, 3]. The space group was reported to be orthorhombic *Pbnm* (or *Pnma* in another setting). It was shown that these compounds have high Néel temperatures $T_N \approx 600 – 700$ K, below which Fe$^{3+}$ moments are ordered antiferromagnetically with a weak ferromagnetic component. With decreasing temperature, the importance of the Fe–*R* interaction increases (certainly for magnetic rare earth ions) leading to the spin-reorientation (SR) transition. This latter takes place at temperature $T_{SR}$, which is often in the range $50 \div 60$ K; though SR transition occurs at much lower temperatures, close to the Neel temperatures of rare-earth alignment for some rare-earth ions like Tb, Yb, or at such a high temperature as ~ 456 K for Sm. The rare-earth subsystem with relatively weak R–R interactions remains paramagnetic at elevated temperatures, or it is weakly polarized by the exchange field of the ordered Fe$^{3+}$ moments. The spontaneous ordering of the rare-earth sublattice takes pace below $T_{NR} \approx 5$–10 K. Complex magnetic properties of the RFeO$_3$ system are governed by the presence of various competing exchange interactions. They include Heisenberg-type super exchange of type Fe–Fe, Fe–R, R–R, and also Dzyaloshinskii-Moriya (DM) interaction [4, 5] which has an important influence on the magnetic properties and leads to weak ferromagnetism.

Recently, the interest in the RFeO$_3$ family of compounds has been greatly renewed because of the discovery of their multiferroic properties. Emergence of ferroelectricity in orthoferrites at temperatures below $T_{NR}$ has been predicted in theoretical work [6] on the basis of a symmetry analysis of the crystal structure. The ordering of the $R^{3+}$ moments takes place in accordance with the Γ1 − Γ8 irreducible representations, which are not compatible with inversion symmetry and thus allow for a linear magnetoelectric effect as well as for a dynamic magnetoelectric effect i.e. an electric-dipole-active magnetic excitations.

In the subsequent experiments on DyFeO$_3$ and GdFeO$_3$ the emergence of a ferroelectric polarization was observed below the magnetic ordering $T_{NR}$ ≈ 5–10 K of the rare-earth subsystem indeed [7, 8]. Later, however, electric polarization in DyFeO$_3$ was found at much higher temperatures, above $T_{SR}$ ≈ 50–60 K [9]. In other orthoferrites like SmFeO$_3$ [10], YFeO$_3$ [11] and LuFeO$_3$ [12], electric polarization was reported even at room temperature. This brings these compounds close to being useful for potential applications in switching elements, sensors, memory and other advanced technical devices with low energy consumption. Macroscopic investigations show a strong influence of external fields on the magnetic and/or ferroelectric properties of these compounds [7, 12, 13, 14].

There is an indication that the DM interaction could be responsible for the emergence of ferroelectric ordering in DyFeO$_3$, YFeO$_3$ and LuFeO$_3$ at high temperatures [12]. It should be noted also that the orthorhombic space group *Pbnm* is centrosymmetric and therefore does not allow spontaneous electric polarization. Therefore, more precise and detailed studies of the crystal and magnetic properties of the RFeO$_3$ compounds required in order to search for the physical origin of the symmetry lowering. Rare-earth orthoferrites RFeO$_3$, with a ferroelectric moment induced presumably by the magnetic DM interaction, are very promising candidates for the realization of such an effect that will also serve to a better understanding of the interplay between ferroelectric polarization and magnetic order.

According to recent precise single-crystal neutron diffraction studies of HoFeO$_3$ below $T_N$ = 647 K Fe sublattice has antiferromagnetic order described by symmetry representation Γ$_4$ [15] with the strongest component along *a*-axis, and weak ferromagnetic component along *c*-axis. The first spin-reorientation phase transition to the antiferromagnetic order with Γ$_1$ takes place at $T_{SR1}$ = 55 K and the second reorientation transition from Γ$_1$ to Γ$_2$ - at $T_{SR2}$ = 35 K, where the strongest component of Fe magnetic moments directed along *c* [15]. Fig. 1 shows the magnetic structures in phases Γ4 and Γ2. Ho order happens at temperatures 3.3 – 4.1 K as it was shown by the works on the heat capacity, magnetization, and Mössbauer studies [16–18]. The magnetic moments of Ho lie in *a-b* plane, the ordering also could be described by Γ2 representation [15] which does not impose any restriction on the moment orientation along c for Ho site 4c.

The spin dynamics of *R*FeO3 orthoferrites have been previously studied with inelastic neutron scattering [19 - 21], Raman spectroscopy [22 - 24], Faraday balance [25 - 27], far-IR [28, 29], and submillimeter [30] spectroscopies. These studies made it possible to determine the parameters of exchange interactions in some orthoferrites, and also discovered a number of unexpected interesting properties of spin dynamics in these materials. Thus the recent study YbFeO$_3$ reveals rich quantum spin dynamics of Yb magnetic sublattice [31]. At temperatures below T$_{SR}$ = 7.6 K the Yb subsystem changes the its excitation spectrum, demonstrating the transition between two regimes with magnon and spinonlike fluctuations. The electromagnon excitations were detected in DyFeO3 and TbFeO3 in magnetic phases, which are compatible with a spontaneous electric polarization [28, 29]. In TbFeO3 was found that the specific exchange of magnons can lead to new magnetic states [32]. Ultrafast control of the spin dynamics by polarized femtosecond laser pulses was observed in DyFeO3 [26] and in TmFeO3 [27].

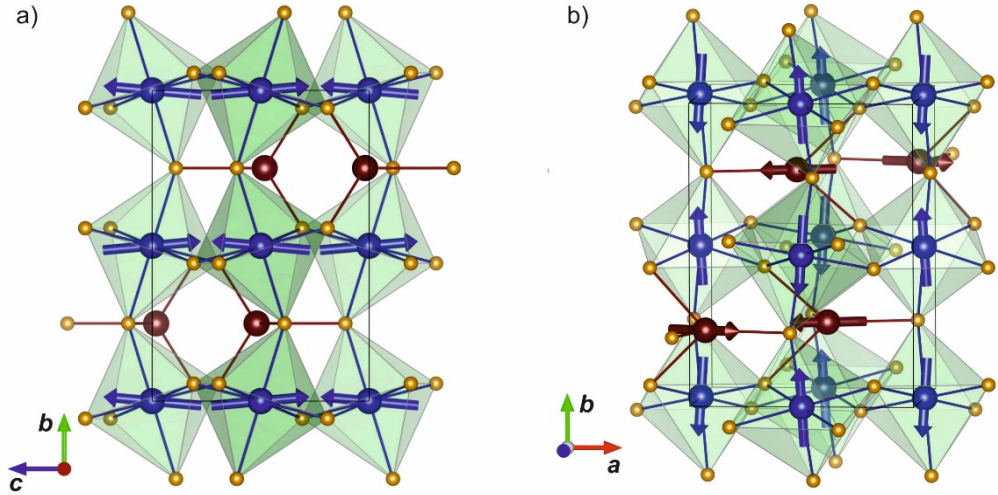

Fig. 1. The magnetic structure: a) in phase Γ4 and b) in phase Γ2. Blue spheres - $Fe^{3+}$ ions, red – $Ho^{3+}$, orange – $O^{2+}$.

Inelastic neutron scattering is the only direct method to investigate the dynamics of a magnetic lattice and, therefore, the most reliable way to obtain the magnetic interaction parameters. This could provide an essential input for the theoretical description of the magnetic properties as well as for developing the model responsible for the multiferroicity. Just a few earlier works were performed on the orthoferrites with Er, Tm [19] and Tb [20] by this technique and the parameters of exchange interactions have been determined, using incomplete models for the calculations. These considered only the strongest exchange interactions of Heisenberg-type like Fe-Fe super exchange interactions between the nearest and next-nearest neighbors, whereas DM antisymmetric exchange almost never have been taken into account. Just recently, the DM interaction and single-ion anisotropy of Fe were considered in the studies of magnetic dynamics of YFeO3 [21] and YbFeO3 [31]. Despite its weakness, the DM-interaction could strongly influence on the magnetic properties of a system. As mentioned above, in the case of RFeO3 orthoferrites, the DM interaction is the origin of weak ferromagnetism and definitely needs to be taken into account. Therefore, we plan to study both the symmetric exchange interactions $3d$–$3d$, $3d$–$4f$, $4f$–$4f$ as well as the antisymmetric interaction in the same magnetic subsystems.

2. **Experimental**

High quality twins-free single crystal of HoFeO3 has been grown using fluxed melt method [33]. The shape of the crystal used in experiment is close to parallelepiped with approximate dimensions 5x4x6 mm$^3$ with the longest dimension along c-axis The parameters of the unit cell were refined at room temperature and at 65 K and appeared to be the same at both temperatures. The obtained structure should be attributed to space group *Pbnm* with cell parameters a = 5.280 Å, b = 5.591 Å, c = 7.602 Å, which corresponds completely to those obtained earlier [2].

The inelastic neutron scattering experiments were performed at ILL on the spectrometers IN12 and IN20. High energy excitations were studied at IN20 – thermal neutron triple axis spectrometer, and low energy range was explored at IN12 – cold neutron triple-axis spectrometer.

The experiment at IN20 was performed at some temperatures corresponding to different magnetic phases: at 65 K – weak ferromagnetic phase $\Gamma_4$, at 35 K – antiferromagnetic phase $\Gamma_1$, at 15 and 2.5 K – weak ferromagnetic phase $\Gamma_2$. During the experiment we used the measurements in "constant-$q$" mode consisted in series of energy scans with sequential steps along $h$ or $l$ directions in the reciprocal space (conditionally $h$-scan or $l$-scan in the following text). For the high energy studies the energy step was taken $\Delta E = 1$ meV along the scan in the energy range 10 – 70 meV. The $q$-step was $\Delta h, \Delta l = 0.2$ rlu. And the measurements were made in the vicinity of node $q = [3\ 0\ 5]$ along $h$ direction in the range from $q = [1\ 0\ 5]$ to $q = [3\ 0\ 5]$ and along $l$ direction from $q = [3\ 0\ 3]$ to $q = [3\ 0\ 5]$.

For low energy transfer the measurements at IN12 were made in the range 0 - 7 meV with the energy step $\Delta E = 0.1$ meV along the scan with the step $\Delta h, \Delta l = 0.2$ rlu. Scans were made in the vicinity of node $q = [1\ 0\ 1]$ along $h$ direction in the range from $q = [0\ 0\ 1]$ to $q = [2\ 0\ 1]$ and along $l$ direction from $q = [1\ 0\ 0]$ to $q = [1\ 0\ 2]$. In this way we obtained a maps of the intensity, reflecting different kinds of inelastic scattering.

For more accurate determination of the energy gap and separation of the tale from elastic peak and inelastic peak intensity at energy transfer range 0 ÷ 1 meV, the scans with better energy resolution $\Delta E = 0.05$ meV were performed in the energy range from 0 to 2.5 meV. In this mode the wave vector of scattered neutrons was kept $k_f = 1.25$ Å$^{-1}$ and the momentum transfer varied from $q = [0.6\ 0\ 1]$ to $q = [1.6\ 0\ 1]$.

In the course of data treatment the positions, intensities and half widths of the measured peaks were calculated using the Winplotr program included in the FullProf Suite [34]. For the INS treatment the preliminary calculations for the description of the Fe subsystem were made using our own code performed in the Wolfram Math environment. All final results are obtained using the SpinW software [35]. For the determination of the uncertainty of the fitted parameters we took the maximum parameter range, at which convergence R-factor did not change its value.

3. **Experimental results**

The typical results of the measurements at high-energy transfer are presented at Fig. 2, where $q$-constant scans for 65K and 2.5K obtained at IN20 are shown. As it can be seen, a significant number of inelastic peaks maxima can be found on the measured spectra.

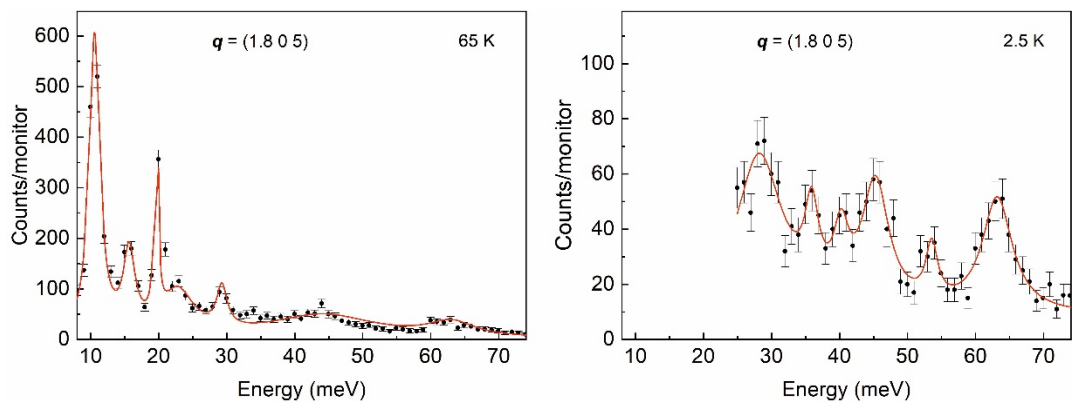

Fig. 2. Typical scans of inelastic neutron scattering, obtained at IN20 at T = 65 K and T = 2.5 K. Solid lines – fitting results.

Measured in this way scans were then combined into energy maps and these ones obtained at IN20 for the higher energy range are presented at Fig. 3. The dispersion branches of $Fe^{3+}$ magnon excitations is clearly visible here. The number of dispersionless branches are also clearly observed on the maps. The energies of these lines correspond to the transitions in molecular field, which was obtained by the estimations within the mean-field approach. The details of this consideration will follow in the separate article soon. Dispersionless branches at ~ 10 meV, ~ 15 meV correspond also to CEF levels reported in the works by optical spectroscopy [36, 37]. It can be seen also that not all these q-dependences are straight lines, that can really be related to some hybridization with magnon or phonon scattering. The low energy transfer results obtained at IN12 are presented at Fig. 4, where the maps along *h* direction around point [1 0 1] from q = [0 0 1] to q = [2 0 1] at temperatures of 65K, 35K and 2.5K are shown. The peak intensities at energy transfer of 0 meV correspond to elastic scattering. At 35 K and 2.5 K one can see doubled dispersionless (or almost dispersionless) lines in the range 0 ÷ 1 meV which should be attributed to split level of the crystal field and Ho-originated magnon branch. At *q* = [1 0 1], dispersion curves with the energy gap are observed, which definitely should be associated with the spin waves in the $Fe^{3+}$ sublattice. In order to elucidate the nature of the signal, we have checked all measured peaks with respect to reasonable resolution widths. For the description of the peaks shape, pseudo-Voight function was used. Analysis shows that different widths of the peaks originate solely from the instrument, excluding any broadening effects from real physics.

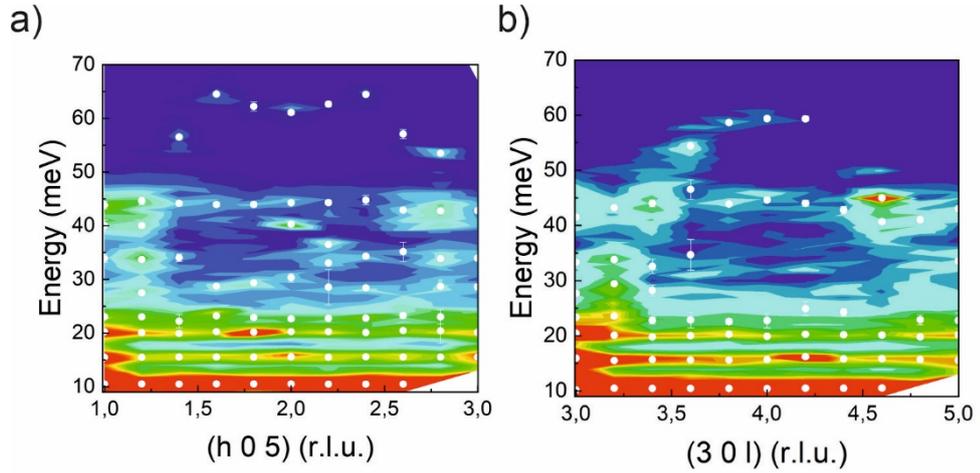

Fig. 3. Energy maps, obtained at IN20 at temperature of 65K: a) *h*-scans with *l* = 5 and b) *l*–scans with *h* = 3. The colors show the intensity, the white dots - positions of the inelastic peaks.

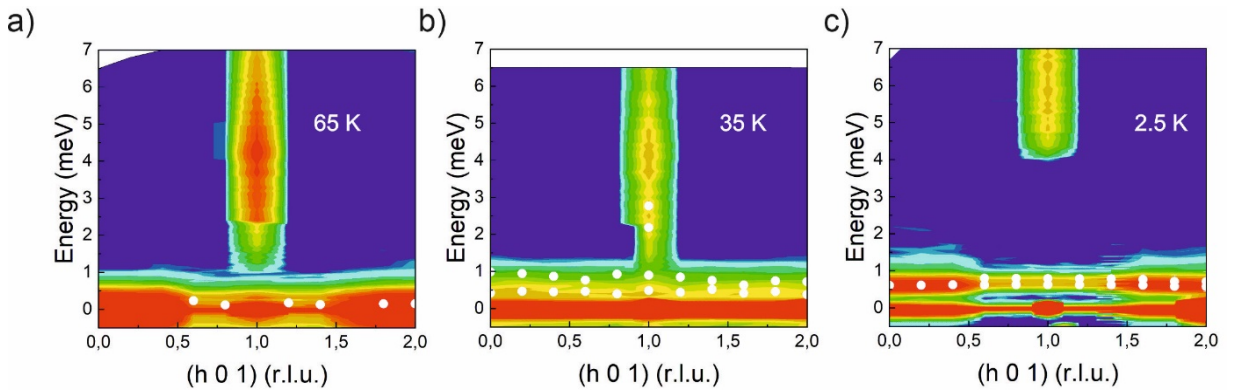

Fig. 4. Energy maps obtained at IN12 by *h*-scans with *l* = 1 at temperatures a) 65 K, b) 35 K and c) 2.5 K. The color indicate the intensity, the white dots are the positions of the inelastic peaks.

4. **Spin waves modeling and calculation**

For the analysis of magnetic excitations observed in HoFeO$_3$ the standard linear spin-wave approach was used. In the general case, a Hamiltonian must contain the following terms:

$$H = H^{Fe-Fe} + H^{Ho-Ho} + H^{Fe-Ho} \qquad (1)$$

where the first two terms describe exchange interactions and single-ion anisotropies within Fe$^{3+}$ and Ho$^{3+}$ subsystems, respectively. The third term describes the interaction between the Fe and Ho subsystems.

At $T = 65$ K subsystem Ho$^{3+}$ supposed to be non-ordered, i.e. has zero ordered magnetic moment and the second and third terms in the Hamiltonian (1) are zero. Then dispersion curves can be described using interactions within the Fe$^{3+}$ sublattice only:

$$H^{Fe-Fe} = \sum_{ij} S_i^{Fe} \cdot J_{ij}^{Fe} \cdot S_j^{Fe} + \sum_i S_i^{Fe} \cdot A_i^{Fe} \cdot S_i^{Fe} + \sum_{mn} S_m^{Fe} \cdot D_{mn}^{Fe} \cdot S_n^{Fe} \qquad (2)$$

where $S$ – spin operator, $J$ – isotropic exchange interactions, $A$ - single ion anisotropy, $D$ - DM interaction parameter. In the expression (2), the parameters of exchange interactions, anisotropy and DM are written as a 3x3 matrix, as it was used in the calculations. In this Hamiltonian (2), the first term dictates an overall shape and maximum energy of the Fe excitations. For the simulation of the exchange interactions in Fe subsystem it is reasonable to include in the analysis the interactions between nearest neighbors ($J_{nn}$) and next-nearest neighbors ($J_{nnn}$). The nearest neighbors exchange interactions along the axis $c$ - $J_c^{Fe}$ and in the plane $ab$ - $J_{ab}^{Fe}$ (see Fig. 5a, b) describe the exchange between ions that are at different distances. In the studies [19 - 21], these exchanges are considered as the equal interactions between the nearest neighbors and are described by one common exchange parameter. However, the latest work [38] on YFeO$_3$ showed that difference in the distance between the ions such a small as ~ 0.03 Å, may result in an appreciable energy difference between the exchange parameters of about ~ 0.4 meV. The similar situation arises when considering the interaction between next-nearest neighbors, where several interactions between ions with close exchange paths distances were described by one common parameter $J_{nnn}^{Fe}$ (Fig. 5 c-e).

The anisotropy determines a magnetic ground state [39] and gives rise to the gap in the Fe magnon spectrum [21]. Due to the orthorhombic symmetry of the Fe$^{3+}$ environment, the anisotropy must be described by two nonequivalent constants $Aab$ and $Ac$. In RFeO$_3$ with nonmagnetic R-ions, a dominating $Aab$ stabilizes the Γ4 phase. In HoFeO$_3$, the Ho-Fe interaction induces renormalization of the effective anisotropy constants. At $T \approx T_{SR1}$, $Aab$ and $Ac$ become approximately equal [39]. Below $T_{SR2}$ $Ac > Aab$ that stabilizes the Γ2 phase.

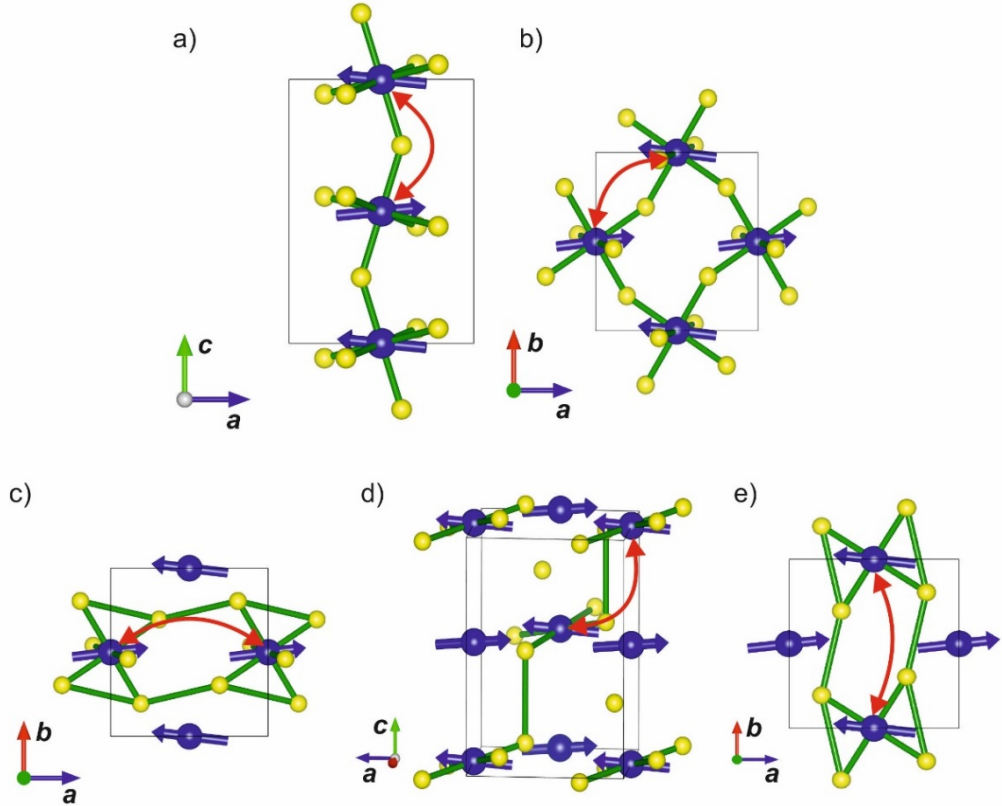

Fig. 5. The schemes of the interaction HoFeO$_3$ under consideration: a) and b) exchange paths of interaction $J_c^{Fe}$ and $J_{ab}^{Fe}$ with distances 3.810A and 3.842A, respectively; c) d) and e) exchange paths of interaction $J_{nnn}^{Fe}$ with distances 5.282A, 5.409A and 5.591A, respectively. Blue spheres - Fe$^{3+}$ ions, orange – O$^{2-}$. Red arrows show the exchange interaction paths.

At temperatures above $T_{SR1} = 55$ K the moments of Fe are directed along $a$ axis and are ordered antiferromagnetically with the propagation vector $\mathbf{k} = (0\ 0\ 0)$. The DM antisymmetric exchange interaction leads to a weak canting of the sublattices, which is described by two constants $D1$ and $D2$, responsible for the canting along $c$ and $b$ axes, respectively. Since DM interaction is very small, for its determination we consider only two pairs of the nearest neighbors in Fe subsystem (Fig. 5a, b). The DM exchange parameters were calculated in the following way: as in [21], initial DM values were obtained based on the canting angles of the sublattices. The values of the canting were taken from [40]. These DM values were used then as starting parameters when fitting our model.

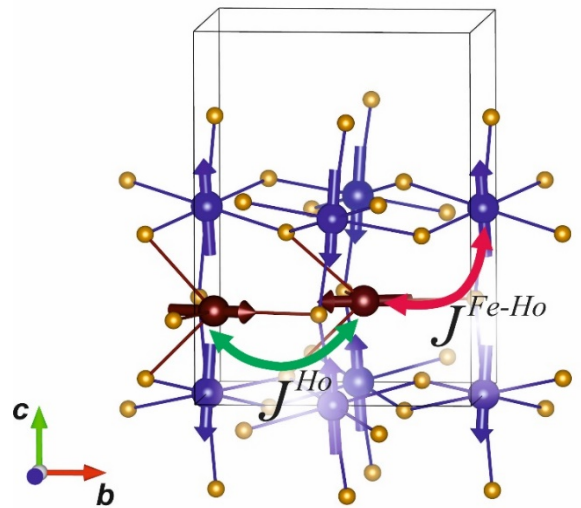

Fig. 6. Blue spheres show Fe ions. Orange – O. Red – Ho ions. Red arrow shows exchange paths $J^{Fe-Ho}$. Green arrow – exchange $J^{Ho}$.

In order to describe the dispersion at lower temperatures at T = 2.5 K, when the moments of $Ho^{3+}$ are ordered, and the system is in the magnetic phase Γ2, it is necessary to take into account exchange interactions between $Fe^{3+}$ and $Ho^{3+}$ subsystems - $J^{Fe-Ho}$, and interactions within $Ho^{3+}$ subsystem $J^{Ho}$ (Fig. 6). Then we the following terms should be added to the expression (1):

$$H^{Fe-Ho} + H^{Ho-Ho} = \sum_{ij} S_i^{Fe} \cdot J_{ij}^{Fe-Ho} \cdot s_j^{Ho} + \sum_{mn} s_m^{Ho} \cdot J_{mn}^{Ho} \cdot s_n^{Ho} \quad (3)$$

where $s^{Ho}$ – the Ho spin moment operator:

$$s^{Ho} = (g_j - 1)J \quad (4)$$

where $g_j$ - Lande factor, $J$ - total angular moment. Since the crystal structure and distances between iron ions have not changed, it is reasonable to fix in fitting procedure the parameters included in $H^{Fe-Fe}$ which were obtained from the data at T = 65 K. For $J^{Ho}$ calculation in order to simplify the model we restrict ourselves to the R-R interaction between the nearest neighbors. This is because the value of exchange interaction within the Ho sublattice is small, and at the same time in our experiment we cannot distinguish the dispersion curves corresponding to Ho.

Experimental and calculated dispersion curves are shown at Fig. 7. Calculated values of all parameters are presented at the Table 1. The obtained values of the exchange parameters inside the Fe sublattice are in good agreement with those ones in other similar compounds [21, 31, 38]. At the same time, the difference between nearest neighbors exchange interactions $J_c^{Fe}$ and $J_{ab}^{Fe}$ is noticeable. With regard to next-nearest neighbors, the results of our calculations show that there is no sense in separation of the parameters because the values of the exchange parameters appeared to be the same within the calculation error.

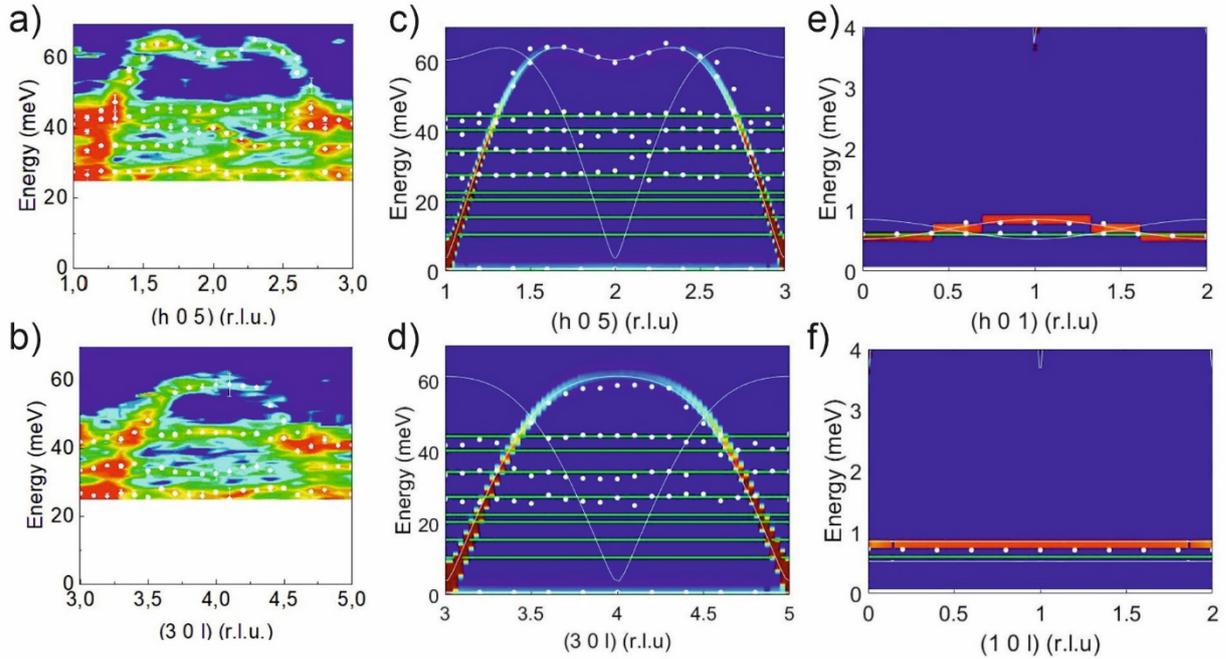

Fig. 7. a), b) High energy magnon dispersion measured at 2.5 K, the colors show the intensity, the white dots - positions of the inelastic peaks. c), d) Calculated high energy dispersion maps, white lines - calculated dispersion curves; green line - levels of the crystal field. e), f) calculated energy maps in the energy range 0 - 4 meV at $T = 2.5$ K; compare with Fig.4.

Table 1. The parameters of exchange interactions (in meV) obtained in this work with the best fit of the data (Rw = 5.43), and similar parameters obtained for another orthoferrites for comparison.

| Magnetic phase Γ4: | $J_c^{Fe}$ | $J_{ab}^{Fe}$ | $J_{nnn}^{Fe}$ | $J^{Fe-Ho}$ | $J^{Ho}$ | D1 | D2 | Ac | Aab |
|---|---|---|---|---|---|---|---|---|---|
| HoFeO$_3$ | 4.901(5) | 4.764(5) | 0.150(7) | | | 0.12(2) | 0.08(2) | 0 | 0.008(1) |
| YFeO$_3$ [21] | 4.77 | | 0.21 | | | 0.074 | 0.028 | 0.003 | 0.0055 |
| YFeO$_3$ [38] | 5.02 | 4.62 | 0.22 | | | 0.1447 | 0.1206 | 0.0025 | 0.0091 |
| YbFeO$_3$ [31] | 4.675 | | 0.158 | | | 0.086 | 0.027 | 0 | 0.033 |
| Magnetic phase Γ2: | | | | | | | | | |
| HoFeO$_3$ | 4.901(5) | 4.764(5) | 0.150(7) | -0.026(2) | 0.035(5) | 0.12(2) | 0.08(2) | 0.017(1) | 0 |
| YbFeO$_3$ [31] | 4.675 | | 0.158 | | | 0.086 | 0.027 | 0.023 | 0 |

## 5. Discussion

The results confirm that the exchange interactions $J_c^{Fe}$ and $J_{ab}^{Fe}$ are the strongest in the system and they are crucial in the formation of structure with antiferromagnetic ordering in HoFeO$_3$, and also these interactions determine the maximum excitation energy of the Fe$^{3+}$ magnetic sublattice.

At high temperatures ordered Fe$^{3+}$ sublattice polarizes Ho$^{3+}$ subsystem, giving rise to Ho-Fe exchange interaction, which in turn, leads to the exchange splitting of the ground state of Ho$^{3+}$. This could be seen on the energy maps at Fig. 4 where the energy level that corresponds to Zeeman splitting of the ground state of Ho lies in the energy range of 0 ÷ 1 meV. At the temperature $T = 65$ K the value of splitting is $\Delta_{HGS} \approx 0.35(1)$ meV. The splitting value increases to $\Delta_{HGS} \approx 0.95(1)$ meV at $T = 35$K, and then begins to decrease due to inset of Ho-Ho interaction, thus lowering the splitting value to $\Delta_{HGS} \approx 0.65(1)$ meV at the $T = 2.5$ K. This gives the evidence that interaction $J^{Fe-Ho}$ and $J^{Ho}$ must be of opposite sign since that provides the observed behavior of the level splitting energy. The similar situation was observed in experiments by optical spectroscopy [36, 37], where the absorption lines splitting was $\Delta_{HGS} \approx 0.25$ meV at T = 100K, at T = 20K it was $\Delta_{HGS} \approx 0.88$ meV and at T = 1.2 K $\Delta_{HGS} \approx 0.61$ meV. The obtained values of $J^{Fe-Ho}$ and $J^{Ho}$ are presented at Table 1. These values and signs correlate well with the results from optical spectroscopy [36, 37], that gives $J^{Fe-Ho}$<0.2 meV, and by the Faraday balance measurements [25] that gives $J^{Fe-Ho} = 0.0215$ meV. In these cases, $J^{Fe-Ho}$ exchange between the Fe and Ho sublattices and the $J^{Ho}$ exchange in Ho sublattice also have opposite signs.

The exchange interaction $J^{Fe-Ho}$ is comparatively small, nevertheless it influences considerably on our model calculations. Since this exchange couples the Fe and Ho subsystems, it contributes to the energy of the dispersion curves of both subsystems. In our experiments, we cannot clearly distinguish the dispersion branches from the Ho sublattice, but we are good at defining the dispersion curves corresponding to the Fe subsystem. Therefore, the influence of the $J^{Fe-Ho}$ exchange value can be determined by the energy of the Fe dispersion curves. For example, for $J^{Fe-Ho} = 0.026$ meV, 0 meV, the calculated value of the energy gap (q=1 0 1) is Δ=3.82 meV, 3.72 meV, respectively. While the calculated maximum energies for corresponding exchange values (at q=1.68 0 1) differ very little: 64.11 meV, and 64.10 meV.

According to our simulations, dispersion curves associated with spin waves in the Ho sublattice should be in the energy range of 0 ÷ 1 meV. In the same energy range, the peaks from the transition in the exchange split crystal field level are present. The resolution of our measurements does not permit to make the perfect separation between the exact positions of peaks corresponding to inelastic scattering by spin waves in Ho sublattice. Nevertheless, the peak splitting is clearly seen at the energy scans at temperatures 35 K and 2.5 K and peak positions could be extracted. As example the energy scans at different temperatures are presented at Fig. 8, where one can see asymmetric peak at the region of 0.7 meV (Fig. 8b and c). This value is close to the position of the split level of the crystal field at low temperatures. Apparently this asymmetric peak broadening is due to the fact that the reflections from excitations of the Ho sublattice lie close to the peaks corresponding to the crystal field levels. At high temperatures, such a splitting is not observed (Fig. 8a). In this way, the calculated value of the Ho-Ho interaction, presented in Table 1, is the maximum possible estimate, on the basis of which the dispersion curves are reproduced in Fig. 7.

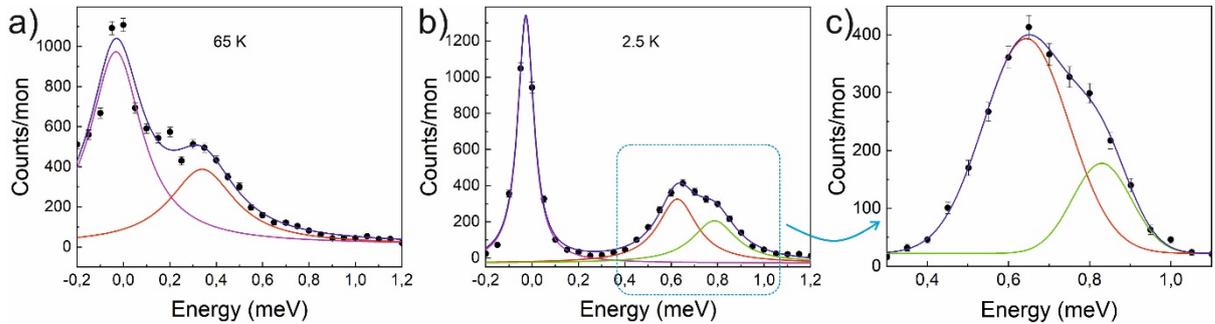

Fig. 8. a), b) – the energy scans in the energy range of -0.2 meV - 1.2 meV at temperatures of 65 K and 2.5 K, respectively. The pink line is the peak corresponding to the elastic scattering of neutrons, the red line - to the level of the crystal field, the green line - to the Ho magnons. Selected region c) – enlarged fragment of the 2.5 K scan.

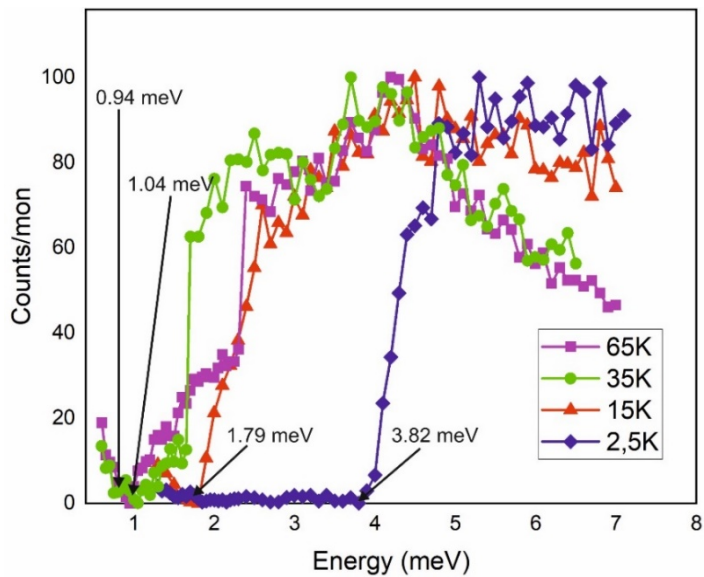

Fig. 9. Energy gaps measured at q = [1 0 1] at different temperatures.

Single-ion anisotropy $A$ leads to the appearance of a gap in the Fe magnon spectrum (Fig. 3). The easy-plane anisotropy $Aab$ dominates in the Γ4 and Γ1 phases, thus forcing iron moments to lie in the $ab$ plane [15]. With temperature decrease, the R-Fe interaction becomes more pronounced and below $T_{SR2}$ it causes a redistribution of energy, and, therefore the change of anisotropy constants. In the Γ2 phase, the easy-axis anisotropy $Ac$ becomes dominant that causes the orientation of iron moments mainly in the direction $c$ [15]. The growth of $Ac$ is confirmed by the experimental fact that the spectrum of magnons with high energies does not change at the spin-orientation transition, while the magnitude of the energy gap changes only. Such behavior can be understood if we assume the temperature dependence of the effective anisotropy constants $Aab\,(T)$ and $Ac\,(T)$. It can be seen at Fig. 9, where the energy gaps are shown at different temperatures. Gap values obtained were $\Delta^{Fe}_{65K}$=0.94 meV and $\Delta^{Fe}_{35K}$=1.04 meV – for the measurements in the Γ4 and Γ1 phases correspondingly, where the easy-plane anisotropy prevails, and their magnitude practically did not change through $T_{SR1}$ transition. At the same time $\Delta^{Fe}_{15K}$=1.79 meV, $\Delta^{Fe}_{2.5K}$=3.82 meV – gaps in the Γ2 phase where easy-axis anisotropy dominates. A sharp increase in the energy gaps could be connected with growth of $Ho^{3+}$ magnetic moment.

## 6. Conclusion

We report the neutron inelastic scattering study of the spin dynamics in $HoFeO_3$ at different temperatures corresponding to three magnetic ordering phases: Γ4, Γ1 and Γ2. The observed spectra were analyzed in the frames of the linear spin-wave theory based on the magnetic structure, derived from the elastic neutron scattering [15]. The values of the parameters of exchange interactions within the Fe-subsystem were obtained, which are in a good agreement with similar values in other orthoferrites. We show that the anisotropy constants $Aab$ and $Ac$ of the iron sublattice in Γ4 and Γ1 phases keep their values, while in the Γ2 phase the ratio between them change itself for the score of growth of $Ac$, thus leading to the increase of anisotropy energy with temperature decrease. This provides the increase of the energy gap in spin-wave spectrum of $Fe^{3+}$ magnetic system. Evaluations of exchange interactions within the Ho-subsystem and between Fe and Ho subsystems were made.

**Acknowledgments**

The authors are grateful to S.V. Maleyev for the discussions on the consideration of CEF. This work was supported by the Russian Foundation for Basic Research grant # 19-52-12047, and DFG grant # SA 3688/1-1. The work of S.N.B. and S.A.G. was done under support of Belorussian Foundation for Basic Research grant #  BRRFFI - NFENK 18К-022.